\documentclass[12pt]{article}
\pdfoutput=1
\usepackage{putex}
\usepackage{graphicx}
\usepackage{caption}
\usepackage{subcaption}
\usepackage{epstopdf}
\usepackage{enumerate}
\usepackage{cite}
\usepackage{tensor}
\usepackage{slashed}
\usepackage[utf8]{inputenc}
\usepackage{mathtools}
\usepackage{amssymb}
\usepackage[]{amsmath}
\usepackage{amscd}
\usepackage{mathrsfs}
\usepackage{bbold}
\definecolor{darkblue}{rgb}{0.1,0.1,.7}
\usepackage[colorlinks, linkcolor=darkblue, citecolor=darkblue, urlcolor=darkblue, linktocpage]{hyperref}
\usepackage[square, comma, compress, numbers]{natbib}

\numberwithin{equation}{section}

\newcommand{\be}{\begin{eqnarray}}
\newcommand{\ee}{\end{eqnarray}}
\newcommand{\eea}{\end{eqnarray}}
\newcommand{\es}[2] {\begin{equation} \label{#1} \begin{split} #2 \end{split} \end{equation}}

\def\Red{\color [rgb]{0.9,0.1,0.1}}

\def\Green{\color [rgb]{0.1,0.6,0.1}}

\def\beq{\begin{equation}} 
\def\eeq{\end{equation}} 
\def\<{\langle}
\def\>{\rangle}

\def\nn{\nonumber} 
 
\def\cO {{\cal O}}

\def\cN {{\cal N}}

\newcommand{\ec}{\,,}

\allowdisplaybreaks
\newcommand\SO{\text{SO}}
\newcommand\Sp{\text{Sp}}
\newcommand\ptl{\partial}
\def\bR {{\bf R}} 
\def\De {\Delta} 
\def\cD {{\cal D}} 
\def\oo {\infty} 
\def\x {\times} 
\def\be#1\ee{\begin{align}#1\end{align}}

\usepackage{braket}
\usepackage{setspace}
\usepackage{array}
\usepackage{titlesec}

\newcolumntype{x}[1]{>{\centering\arraybackslash\hspace{0pt}}p{#1}}

\usepackage{lipsum}

\begin{document}

\preprint{PUPT-2490}

\institution{PU}{Joseph Henry Laboratories, Princeton University, Princeton, NJ 08544}
\institution{Yale}{Department of Physics, Yale University, New Haven, CT 06520}
\institution{IAS}{School of Natural Sciences, Institute for Advanced Study, Princeton, New Jersey 08540}

\title{Fermion-Scalar Conformal Blocks}

\authors{Luca Iliesiu,\worksat{\PU} Filip Kos,\worksat{\Yale} David Poland,\worksat{\Yale, \IAS}\\[10pt] Silviu S.~Pufu,\worksat{\PU} David Simmons-Duffin,\worksat{\IAS} and Ran Yacoby\worksat{\PU}}

\abstract{We compute the conformal blocks associated with scalar-scalar-fermion-fermion 4-point functions in 3D CFTs. Together with the known scalar conformal blocks, our result completes the task of determining the so-called `seed blocks' in three dimensions. Conformal blocks associated with 4-point functions of operators with arbitrary spins can now be determined from these seed blocks by using known differential operators.}

\date{}

\maketitle

\newpage

\tableofcontents

\newpage

\section{Introduction}
\label{sec:intro}

The conformal bootstrap \cite{Ferrara:1973yt,Polyakov:1974gs,Mack:1975jr} has recently resurfaced as a powerful numerical tool for constraining the algebra of local operators of conformal field theories (CFTs) in $D > 2$ spacetime dimensions \cite{Rattazzi:2008pe, Rychkov:2009ij, Caracciolo:2009bx, Poland:2010wg, Rattazzi:2010gj, Rattazzi:2010yc, Vichi:2011ux, Poland:2011ey, Rychkov:2011et, ElShowk:2012ht,Liendo:2012hy, Beem:2013qxa, Kos:2013tga, El-Showk:2013nia, Alday:2013opa, Gaiotto:2013nva,Bashkirov:2013vya, Berkooz:2014yda, El-Showk:2014dwa, Nakayama:2014lva,Nakayama:2014yia, Alday:2014qfa, Chester:2014fya, Kos:2014bka, Caracciolo:2014cxa, Nakayama:2014sba, Golden:2014oqa, Chester:2014mea, Paulos:2014vya, Beem:2014zpa, Simmons-Duffin:2015qma, bobev2015bootstrapping, bobev2015bootstrapping2, Kos:2015mba, chester2015accidental, Beem:2015aoa, iliesiu2015bootstrapping,poland2015exploring,Lemos:2015awa}. After achieving a great degree of success in constraining  four-point functions of scalar operators in various dimensions, in \cite{iliesiu2015bootstrapping} the numerical bootstrap was first applied to four-point functions of spin-$1/2$ fermions in $3$D CFTs. The preliminary numerical study of \cite{iliesiu2015bootstrapping} suggests that the fermionic bootstrap may give access to CFTs that are otherwise hard to study by analyzing only correlators of scalars, such as the Gross-Neveu-Yukawa models or the $\cN=1$ supersymmetric Ising model.

One natural way to push this numerical program further is to study the bootstrap constraints associated with a mixed system of four-point functions of scalars and spin-$1/2$ fermions. Indeed, so far the numerical bootstrap has been used to place constraints on operators that appear in the operator product expansions (OPEs) of either two scalars or of two fermions.  All the operators  in these OPEs are  bosonic. In contrast, the mixed scalar-fermion bootstrap would allow us to also place constraints on the fermionic operators that appear in the OPE of a scalar with a fermion, such as the supercurrent in a supersymmetric theory.  The goal of this paper is to compute the conformal blocks that are necessary in order to apply the bootstrap to such a mixed system of correlators.

In general, the main theoretical obstruction to applying the bootstrap to four-point functions of operators with spin is that the associated conformal blocks are not always known. However, there has been some progress in our understanding of spinning conformal blocks since the original work \cite{DO1,DO2,DO3} of Dolan and Osborn on conformal blocks of scalar four-point functions. In particular, at least in some cases it is now known that spinning conformal blocks can be obtained by acting with certain differential operators on the conformal blocks of a distinguished subset of four-point functions, which are known as ``seed correlators.'' It is straightforward to determine these differential operators  \cite{Costa:2011dw,echeverri2015deconstructing,iliesiu2015bootstrapping}, but the seed blocks are not known in general.\footnote{For examples of non-trivial seed blocks that are known, see \cite{SimmonsDuffin:2012uy,Costa:2014rya,rejon2015scalar}.} 

In $D\geq 4$ the full set of seed correlators is infinite, but in $3$D, because of the simplicity of the Lorentz group, this set is finite. In fact, in $3$D there are only two seed four-point functions: one with four scalars, and the other with two scalars and two spin-$1/2$ fermions. The blocks of the first seed correlator are well-known~\cite{ElShowk:2012ht, Hogervorst:2013sma, Kos:2013tga, Kos:2014bka}, and can be used to determine the blocks corresponding to any four-point function of four bosonic or four fermionic operators using the differential operators constructed in \cite{Costa:2011dw,iliesiu2015bootstrapping}. The blocks of the second seed correlator are of two types.  The first type corresponds to an exchange of a bosonic operator that appears both in the OPE of the two scalars and in the OPE of the two fermions;  these blocks, too, can be obtained from the scalar conformal blocks using the differential operators in \cite{Costa:2011dw,iliesiu2015bootstrapping}.  The second type of blocks corresponds to an exchange of a fermionic operator that appears in the OPE of a scalar with a fermion.  These blocks are not currently known and will be determined in this paper.  Using the same differential operators as  in \cite{Costa:2011dw,iliesiu2015bootstrapping}, they can be used to further determine the blocks corresponding to a fermionic exchange in a four-point function of two bosonic and two fermionic operators of arbitrary spin.

As done in \cite{Kos:2014bka, Kos:2013tga} for scalar blocks, we will determine a recursion formula expression for our fermion-scalar blocks. We start by using the embedding space formalism, following the conventions presented in \cite{iliesiu2015bootstrapping}, to constrain the generic form of the three-point function $\< \Psi(x_1)\Phi(x_2) \cO_{\ell}(x)\>$ (Section~\ref{threePoint}) and of the four-point function $\langle {\Psi_1(x_1) {\Phi_2(x_2)\Phi_3} (x_3)\Psi_4}(x_4)\rangle$ (Section \ref{fourPoint}). In Sections~\ref{recursionSection}--\ref{selRules}, we obtain an ansatz for a recursion formula of the conformal blocks. In Section~\ref{results}, we iteratively solve each order of the recursion using the Casimir equations presented in Appendix \ref{AppendixCasEq}. We end with a brief discussion in Section~\ref{DISCUSSION} that includes the decomposition of a free theory four-point function in terms of our blocks.  Finally, in Appendix~\ref{AppendixRelation3Pt4Pt}, we relate the coefficients that appear in the conformal block expansion to products of OPE coefficients as defined through our normalization conventions. 

\section{Mixed Fermion-Scalar Conformal Blocks}
\subsection{Embedding of Spinor Fields}
\label{embeddingSection}

In this section we set up our conventions for embedding 3D spinor fields into a 5D space by following \cite{iliesiu2015bootstrapping}.
We work in flat 3D spacetime with metric $\eta^{\mu\nu} = \textrm{diag}(-1,1,1)$ and coordinates $x^\mu$ ($\mu,\nu = 0, 1, 2$). The 3D conformal group $\SO(3,2)$ acts linearly in 5D embedding space with metric $\eta^{AB} = \eta_{AB} = \textrm{diag}(-1,1,1,1,-1)$ and coordinates $X^A$ ($A, B=0,\ldots,4$). We will use light-cone coordinates $X^\pm = X^4 \pm X^3$, and write the components of $X$ from now on as $X = (X^\mu, X^+, X^-)$.

The 3D spacetime is embedded in 5D as the projective null cone, defined to be the set of points satisfying $X\cdot X=\eta_{AB}X^A X^B =0$ and identified by $X^A \sim \lambda X^A$. The exact relation between $x^\mu$ and  $X^A$ can be written as
\be
\label{eq:projectivelightcone}
x^\mu = \frac{X^\mu}{X^+}\,, \qquad X=  X^+(  x^\mu ,1 ,  x^2 )\,,
\ee
where $x^2 \equiv \eta_{\mu\nu} x^\mu x^\nu$. 

Let $\cO^{\alpha_1\cdots\alpha_{2\ell}}(x)$ be a dimension $\Delta$ primary operator in the spin-$\ell$ irrep of the  $\mathrm{Spin}(2,1)\simeq \Sp(2,\bR)$ double-cover of the 3D Lorentz group.\footnote{$\Sp(2,\bR)$ indices $\alpha,\beta=1,2$ are raised and lowered with the symplectic form $\Omega_{\alpha\beta}=\Omega^{\alpha\beta} = \begin{psmallmatrix}
0 & 1 \\ -1 & 0
\end{psmallmatrix}$ and are suppressed in contractions using the convention $\psi\chi\equiv \psi_{\alpha}\chi^{\alpha}$. We turn vectors into bi-spinors using $x^{\alpha}{}_{\beta}\equiv x^{\mu}(\gamma_{\mu})^{\alpha}{}_{\beta}$, where $\gamma^{\mu}\equiv (i\sigma^2,\sigma^1,\sigma^3)$. Similarly, embedding $\Sp(4,\bR)$ spinor indices, $I,J=1,\ldots,4$, are raised and lowered with $\Omega_{IJ}=\Omega^{IJ} = \begin{psmallmatrix}
0 & \mathbb{1} \\ -\mathbb{1} & 0
\end{psmallmatrix}$, and contracted as $\Psi\Lambda\equiv \Psi_{I}\Lambda^{I}$. Embedding vectors $X^A$ are written as embedding bi-spinors defined by $X^I{}_J\equiv X^A (\Gamma_A)^I{}_J$, where $\Gamma^A\equiv (\gamma^2\otimes\gamma^0 \ec  \mathbb{1}\otimes\gamma^1 \ec \mathbb{1}\otimes\gamma^2 \ec \gamma^0\otimes\gamma^0 \ec \gamma^1\otimes\gamma^0)$.} To suppress spinor indices on operators we introduce commuting polarizations $s_{\alpha}$ and define $\cO_{\ell}(x,s)\equiv s_{\alpha_1}\cdots s_{\alpha_{2\ell}}\cO^{\alpha_1\cdots\alpha_{2\ell}}(x)$. The embedding space parent of $\cO_{\ell}(x,s)$ is given by
\begin{align}
\cO_{\ell}(X,S) &\equiv S_{I_1}\cdots S_{I_{2\ell}}\cO^{I_1\cdots I_{2\ell}}(X) \ec
\end{align}
where $S_I$ are embedding space spinor polarizations transforming in the fundamental of $\Sp(4,\bR)\simeq\mathrm{Spin}(3,2)$. The operator $\cO_{\ell}(X,S)$ has the following homogeneity properties:
\begin{align}
\cO_{\ell}(aX,bS) &= a^{-(\Delta+\ell)} b^{2\ell}\cO_{\ell}(X,S) \,. \label{Oscaling}
\end{align}

The projection $\cO_{\ell}(X,S)\to \cO_{\ell}(x,s)$ is given by the prescription
\begin{align}
\cO_{\ell}(X,S) = \frac{1}{(X^+)^{\Delta}} \cO_{\ell}(x,s) \ec
\label{eq:projectdowntoflatspace}
\end{align}
where $X$ is related to $x$ according to \eqref{eq:projectivelightcone}, and $S$ is related to $s$ through
  \es{STos}{
  S_I = \sqrt{X^+} \begin{pmatrix}
   s_\alpha \\
   -x^\beta{}_\gamma s^\gamma 
  \end{pmatrix} \,.
  }
Note that $S_I$ satisfies the transversality condition 
 \be
   S_I X^I{}_J = 0 \,.  \label{Transversality}
 \ee
 
The form of correlation functions in embedding space is constrained by the $\SO(3,2)$ symmetry (which acts linearly on the embedding coordinates), homogeneity $\eqref{Oscaling}$, and transversality \eqref{Transversality}.
 
\subsection{The Three-Point Function }
\label{threePoint}

Consider the 3-point function of a spin-$\frac{1}{2}$ fermion primary $\Psi(X_1, S_1)$ of dimension $\Delta_1$, a scalar $\Phi(X_2)$ of dimension $\Delta_2$, and a third fermion $\cO_{\ell}(X_3,S_3)$ of half-odd integer spin $\ell$ and dimension $\Delta_3$. Conformal invariance restricts the form of this 3-point function to be
\begin{align}
\langle \Psi(X_1, S_1)\Phi(X_2)  \cO_{\De_3,\ell}(X_3, S_3)\rangle = \frac{ \lambda^+_{\Psi \Phi \cO} \, r_+ +\lambda^-_{\Psi \Phi \cO} \, r_- }{X_{12}^{\frac{\Delta_{1} + \Delta_2 - \Delta_3 - \ell + 1/2}2}X_{23}^{\frac{\Delta_2 + \Delta_3 - \Delta_{1}+ \ell - 1/2}2}X_{31}^{\frac{\Delta_3 + \Delta_{1} - \Delta_2+\ell + 1/2}2}}\,,
\label{eq:threePointFunc}
\end{align}
where $X_{ij} \equiv -2 X_i \cdot X_j$, $\lambda_{\Psi\Phi\cO}^{\pm}$ are independent OPE coefficients, and where the structures $r_\pm$ can be chosen such that
\be
\label{parityEven3Pt}
r_+ &= \frac{\Braket{S_1 S_3}\Braket{S_3 X_1 X_2 S_3}^{\ell-\frac{1}2}}{X_{12}^{\ell-\frac{1}2}}\,, \\\label{parityOdd3Pt}
r_- &= \frac{X_{13}^{1/2}\Braket{S_1X_2 S_3}\Braket{S_3 X_1 X_2 S_3}^{
\ell-\frac{1}2}}{ X_{23}^{1/2} X_{12}^\ell} \,.
\ee
As in \cite{iliesiu2015bootstrapping}, the angle-brackets are defined by
\be
\<S_1 X_2 X_3 \dots S_n \> \equiv S_{1I} {{X_2}^I}_J {{X_3}^J}_K \dots \Omega^{LM} S_{nM}\,.
\ee

The function multiplying $\lambda_{\Psi\Phi\cO}^+$ is even under parity, $X_k \rightarrow -X_k$, while the one multiplying $\lambda_{\Psi\Phi\cO}^-$ is odd. Throughout the rest of this paper, we will use $\pm$ indices to track the parity of the 3-point structures, of the corresponding OPE coefficients, or of the conformal blocks. Furthermore, from unitarity, if the operators $\Psi(X_1, S_1)$ and $\cO_{\De_3,\ell}(X_3, S_3)$ are real, then Fermi statistics requires that the OPE coefficients $\lambda^\pm_{\Psi\Phi\cO}$ are pure imaginary.

Projecting down to $\bR^{2, 1}$ from embedding space, we can write
\begin{align}
\<\Psi_1(x_1,s_1)\Phi_2(x_2)\cO_{\De,\ell}(x_3,s_3)\> &= \lambda_{\Psi\Phi\cO}^+ \mathcal{R}^+_{\De,\ell}+\lambda_{\Psi\Phi\cO}^- \mathcal{R}^-_{\De,\ell} \,,
\label{eq:threeptstructures}
\end{align}
where the structures $\mathcal{R}^\pm_{\De,\ell}$ are obtained by applying the procedure described in Section~\ref{embeddingSection} to the structures in (\ref{eq:threePointFunc}). We explicitly write $\mathcal{R}^\pm_{\De,\ell}$ in Appendix \ref{AppendixRelation3Pt4Pt}. The $\mathcal{R}^\pm_{\De,\ell}$ depend on all the quantities $\De_1,\De_2,\De,\ell$, but we have chosen to only highlight their dependence on $\De,\ell$ in order to avoid cluttering.

\subsection{The Four-Point Function}
\label{fourPoint}

Let $\Psi_{1,4}$ be spin-$\frac{1}{2}$ fermions of dimension $\Delta_{1,4}$, and $\Phi_{2,3}$ be dimension $\Delta_{2,3}$ scalars. The 4-point function of these operators can be written as
 \es{mixedGeneralForm}{
   \langle  \Psi_1(X_1, S_1)\Phi_2(X_2) \Phi_3 (X_3)  \Psi_4(X_4, S_4)\rangle 
    &= \left(\dfrac{X_{24}}{X_{14}} \right)^{
     \frac{\Delta_{12}}2} \left(\dfrac{X_{13}}{X_{14}} \right)^{
     \frac{\Delta_{43}}2}
     \frac{\sum_I t_I \,  g^I(u,v)}{X_{12}^{\frac{\Delta_1+\Delta_2}2} 
     X_{34}^{\frac{\Delta_3+\Delta_4}2}} \,, 
 }
where $u$ and $v$ are conformally-invariant cross ratios defined as
 \es{uvDef}{
u \equiv \frac{X_{12}X_{34}}{X_{13}X_{24}}\,, \qquad v \equiv \frac{X_{32}X_{14}}{X_{13}X_{24}}\,,
 }
and where the $t_I$ ($I=1,\ldots ,4$) form a basis for the four independent tensor structures  allowed by conformal invariance.  We choose our basis to consist of the two parity even 4-point structures 
 \be
 \label{eq:parityEvenBasis}
t_1 =  i \frac{\Braket{S_1 S_4}}{\sqrt{X_{14}}} \,,\qquad
t_2 = -  i \frac{\Braket{S_1 X_2 X_3 S_4}}{\sqrt{X_{12} X_{34} X_{23}}} \,,
 \ee
and the two parity odd 4-point structures
 \be
 \label{eq:parityOddBasis}
 t_3 =-  i \frac{\Braket{S_1 X_2 S_4}}{\sqrt{X_{12} X_{24} }}\,,\qquad
t_4 = - i \frac{\Braket{S_1 X_3 S_4}}{\sqrt{X_{13} X_{34}}} \,.
 \ee

It will sometimes be convenient to use the coordinates $(r, \eta)$ introduced in \cite{Hogervorst:2013sma} instead of $(u, v)$ in \eqref{uvDef}.  The relation between them is
 \es{uvToreta}{
u = \frac{16r^2}{(1+r^2+2r\eta)^2} \,,  \qquad  v = \frac{(1+r^2-2r\eta)^2}{(1+r^2+2r\eta)^2} \,.
 }

\subsection{Recursion Relations}
\label{recursionSection}

The functions $g^I(u,v)$ can be expanded in conformal blocks as follows
\be
\label{eq:blockgI}
g^I(u,v) &= \sum_{\cO} \sum_{a,b=\pm} \lambda^a_{\Psi_1\Phi_2\cO}\lambda^b_{\Psi_4\Phi_3\cO} g^{I,ab}_{\De,\ell}(u,v).
\ee
Here, $\De,\ell$ label the dimension and spin of the exchanged operator $\cO$.  Note that $\ell$ is a half-odd integer in the present case. The indices $a,b$ label three-point tensor structures (\ref{eq:threeptstructures}). Furthermore, as described in Appendix \ref{AppendixRelation3Pt4Pt}, the OPE coefficients $\lambda^a_{\Psi_i\Phi_j \cO}$ are the same as those appearing in the three-point function (\ref{eq:threePointFunc}) for an appropriate normalization of the two point function of $\cO$. 

The conformal blocks are defined in terms of a sum over states in an individual conformal multiplet as
\begin{align}
\label{eq:BosonFermion}
\frac{1}{X_{12}^{\frac{\Delta_{1}+\Delta_{2}}2} X_{34}^{\frac{\Delta_{3}+\Delta_{4}}2}} &\left(\dfrac{X_{24}}{X_{14}} \right)^{\frac{\Delta_{12}}2} \left(\dfrac{X_{13}}{X_{14}} \right)^{\frac{\Delta_{43}}2} \sum_{I,a,b}  \lambda^{a}_{\Psi_1\Phi_2\cO} \lambda^b_{\Psi_4\Phi_3\cO} t_I g^{I,ab}_{\De,\ell}(u, v)\nn
\\
&= \sum_{\alpha = \mathcal{O}, P\mathcal{O}, \ldots}  \frac{\< \Psi_1(X_1, S_1)\Phi_2(X_2)|\alpha\>\<\alpha| \Phi_3(X_3)  \Psi_4(x_4, S_4) \>}{\<\alpha|\alpha\>},
\end{align}
where $\alpha$ runs over primaries $|\alpha\>=|\cO\>$ and descendants $|\alpha\>=P|\cO\>,\dots$.

This expression can be understood in terms of its analytic properties in $\De$.  Here, we follow the discussion and notation of \cite{penedones2015recursion}.  Poles in $\De$ occur when one of the descendant states $|\alpha\>$ becomes a primary. Indeed, this is only possible if $|\alpha\>$ is null (otherwise acting on $|\alpha\>$ with a special conformal generator would produce a state with nonvanishing norm), so that the corresponding denominator on the right-hand side of \eqref{eq:BosonFermion} vanishes. Descendants of $|\alpha\>$ are then also null states and each gives a contribution to the residue of the pole.

Specifically, there exist a differential operator $\cD_A$ such that the descendant state
\be
|\cO_A,s\> &\equiv \cD_A\cO(x,s)|0\>
\ee
becomes a null primary when $\De\to\De^\star_A$,
\be
\label{eq:twoptresidue}
\<\cO_A,s|\cO_A,s'\> = (\De-\De^\star_A)Q_A^{-1} \mathcal{I_{\ell_A}}(s,s') + O((\De-\De^\star_A)^2) \,.
\ee
The operator $\cO_A$ has spin $\ell_A$ and dimension $\De+n_A$, where $n_A \geq 1$ is the degree of $\cD_A$ in $\partial_x$.
Here, $\mathcal{I_{\ell_A}}$ is a canonical tensor structure for a two-point function of operators with spin $\ell_A$.  The constant $Q_A$ depends on the differential operator $\cD_A$.

When $\De=\De_A^\star$, the operator $\cO_A$ transforms like a primary with dimension $\De_A\equiv \De_A^\star+n_A$ and spin $\ell_A$.  Thus, its three-point function with $\Psi\Phi$ can be expressed in terms of our basis of three-point structures (\ref{eq:threeptstructures}),
\be
\label{eq:threeptmatrix}
\cD_A \mathcal{R}^a_{\De,\ell} \Bigr|_{\Delta = \Delta_A^\star} &= (M_A)^a_b \mathcal{R}^b_{\De_A,\ell_A}  \,.
\ee
where $\mathcal{R}^b_{\De_A,\ell_A}$ depends on the scaling dimensions of both the fermionic ($\Psi$) and scalar ($\Phi$) operators, while $M_A$ depends on the difference in scaling dimensions, $\Delta_{\Phi} - \Delta_{\Psi}$.

The residue of a conformal block at $\De=\De_A^\star$ comes from the contribution of $\cO_A$ and all of its descendants (which also become null as $\De\to \De_A^\star$).  Thus, it is proportional to a conformal block for $\cO_A$.  The constant of proportionality follows from (\ref{eq:twoptresidue}) and (\ref{eq:threeptmatrix}),
\be
\label{eq:simplepoleinblock}
g_{\De,\ell}^{I,ab} =  \frac{(M_A^{(L)})^a_c Q_A (M_A^{(R)})^b_d}{\De-\De_A^\star} g_{\De_A,\ell_A}^{I,cd} + O((\De-\De_A^\star)^0)\,,
\ee
where the superscripts $L$ or $R$ indicate that we compute $M_A$ using the left three-point function $\<\Psi_1\Phi_2\cO\>$ or the right three-point function $\<\Psi_4\Phi_3\cO\>$. In our case, the left and right three-point functions involve the same kinds of operators, so the $M_A$ are simply related by $\Psi_1 \Phi_2 \leftrightarrow \Psi_4 \Phi_3$. Since $M_A$ only depends on the difference between the dimension of the scalar and fermion operator, we can write
\be
M^{(L)}_A = M^{\De_{12}}_A\,, \qquad
M^{(R)}_A = M^{\De_{43}}_A\,.
\ee

In the case studied here, the only poles in the conformal blocks are simple poles of the form (\ref{eq:simplepoleinblock}).\footnote{See \cite{penedones2015recursion} for a detailed discussion of what kind of poles can occur in more general situations.}  Furthermore, after stripping off a factor $r^\De$, the blocks have a finite limit as $\De\to\oo$.  It follows that the blocks satisfy a recursion relation of the form
 \es{Recursion}{
g_{\De,\ell}^{I,ab}(r,\eta) &= r^\De h_{\De,\ell}^{I,ab}(r,\eta) \,, \\
h_{\De,\ell}^{I,ab}(r,\eta) &= h_{\oo,\ell}^{I,ab}(r,\eta) + \sum_A r^{n_A} \frac{(M_A^{(L)})^a_c Q_A (M_A^{(R)})^b_d}{\De-\De_A^\star} h_{\De_A,\ell_A}^{I,cd}(r,\eta) \,.
 }
(See \eqref{uvToreta} for the relation between $(r, \eta)$ and the cross-ratios $(u, v)$ defined in \eqref{uvDef}.)

\subsection{Selection Rules}
\label{selRules}

Because we have chosen our tensor structures to have definite parity, the indices $I,a,b$ satisfy selection rules, regardless of whether the CFT we are studying is invariant under parity.  A parity even four-point structure $I_+=1,2$ can only arise from a combination of two parity even or two parity odd three-point structures ($ab=++,--$).  Similarly, a parity odd four-point structure $I_-=3,4$ can only arise from a combination of three-point structures of opposite parity ($ab=+-,-+$).  Thus, the only nonzero blocks are
\be
g^{I_+,++}_{\De,\ell},\quad g^{I_+,--}_{\De,\ell}, \quad
g^{I_-,-+}_{\De,\ell}, \quad g^{I_-,+-}_{\De,\ell},\qquad (I_+=1,2,\ I_-=3,4).
\ee

The coefficients $(M_A)^a_b$ obey additional selection rules depending on the parity properties of the differential operator $\cD_A$:
\begin{itemize}
\item If $\cD_A$ preserves parity, then only $(M_A)^+_+$, $(M_A)^-_-$ are nonzero.
\item If $\cD_A$ changes parity, then only $(M_A)^+_-$, $(M_A)^-_+$ are nonzero.

\end{itemize}

For convenience, let us group the nonzero blocks into $2$-vectors.
\be
\label{eq:crazyrelabeling}
{\mathbf{g}}_{\De,\ell}^{I_+} \equiv \begin{pmatrix}g^{I_+,++}_{\De,\ell} \\ g^{I_+,--}_{\De,\ell}\end{pmatrix},\qquad
{\mathbf{g}}_{\De,\ell}^{I_-} \equiv \begin{pmatrix}g^{I_-,+-}_{\De,\ell} \\ g^{I_-,-+}_{\De,\ell}\end{pmatrix}.
\ee
Our recursion relation then has the form
\begin{equation}
\label{eq:recurssionFormula}
{\mathbf{h}}_{\Delta, \ell}^{I_\pm}(r, \eta) = {\mathbf{h}}^{I_{\pm}}_{\infty,\ell}(r, \eta) + \sum_A \dfrac{ r^{n_A}}{\Delta-\Delta_A^\star} \mathbf c_A^\pm\,{\mathbf{h}}_{\Delta_A, \ell_A}^{I_\pm}(r, \eta),
\end{equation}
where ${\mathbf{h}}_{\De,\ell}^I$ is defined analogously to (\ref{eq:crazyrelabeling}), and the $2\x2$ matrices $\mathbf{c}_A^\pm$ are
\be
\label{eq:cmatrices}
\mathbf{c}_A^+ &\equiv Q_A
\begin{pmatrix}
(M_A^{\De_{12}})^+_+ (M_A^{\De_{43}})^+_+
&
(M_A^{\De_{12}})^+_- (M_A^{\De_{43}})^+_- 
\\
(M_A^{\De_{12}})^-_+ (M_A^{\De_{43}})^-_+
&
(M_A^{\De_{12}})^-_- (M_A^{\De_{43}})^-_-
\end{pmatrix},\nn\\
\mathbf{c}_A^- &\equiv Q_A
\begin{pmatrix}
(M_A^{\De_{12}})^+_+ (M_A^{\De_{43}})^-_-
&
(M_A^{\De_{12}})^+_- (M_A^{\De_{43}})^-_+
\\
(M_A^{\De_{12}})^-_+ (M_A^{\De_{43}})^+_-
& 
(M_A^{\De_{12}})^-_- (M_A^{\De_{43}})^+_+
\end{pmatrix}.
\ee
If $\cD_A$ preserves parity then $\mathbf{c}_A^\pm$ will be diagonal, while if $\cD_A$ changes parity then $\mathbf{c}_A^\pm$ will be anti-diagonal.

\section{Results}
\label{results}

Since each operator in the four-point function (\ref{eq:blockgI}) is an eigenvector of the Casimir operator of the conformal group, the functions $g^{I} (r, \eta)$ will have to satisfy a set of Casimir equations which we derive in Appendix~\ref{AppendixCasEq}. Due to our choice of structures, the Casimir equations decouple into two sets of equations \eqref{CasEqs}, one corresponding to parity even blocks and one to parity odd blocks.

Our strategy for computing the blocks $g^{I, ab}(r, \eta)$ is as follows: using the recursion ansatz obtained from the state-operator correspondence (\ref{eq:recurssionFormula}), which we plug into \eqref{CasEqs}, we  firstly solve for $\mathbf h^I_{\infty, \ell}(r, \eta)$ and then compute the residue matrices $\mathbf c_A^\pm$ to all orders in $r$.  

\subsection{Blocks in the Limit $\De\to \oo$}
\label{parityEvenSection}

As explained in Appendix~\ref{AppendixCasEq}, the functions ${\mathbf{h}}^{I}_{\oo,\ell}$ can be determined by solving the Casimir equation in the limit $\Delta \rightarrow \infty$.  The equation is automatically satisfied at order $\De^2$.  The $r$-dependence of the solution is fixed by solving the equation at order $\De^1$.  The $\eta$-dependence can be determined by examining the equation at order $\De^0$ and small $r$.  This leaves two linearly independent solutions for $\mathbf{h}^{1,2}_{\oo,\ell}$ and two linearly independent solutions for $\mathbf{h}^{3,4}_{\oo,\ell}$.  We fix the correct linear combinations by matching to the OPE limit $r\to 0,v\to 1$ in Appendix~\ref{AppendixRelation3Pt4Pt}.  The result is
 \begin{align}
 \label{h1Def}
  {\mathbf{h}}^{1}_{\oo,\ell}(r, \eta) &= 
   \frac{ v^{\frac{1}4(\Delta_{12}+\Delta_{43})} }
             { \sqrt{1-r^2} \sqrt{1 + r^2 + 2 r \eta}} 
     \begin{pmatrix}
       \frac{ -P_{\ell-1/2}^{(0, 1)}(\eta)}{1-r}  -\frac{P_{\ell-1/2}^{(1, 0)}(\eta)}{1+r}  \\[5pt] 
         \frac{- P_{\ell-1/2}^{(0, 1)}(\eta)}{1-r}   +  \frac{ P_{\ell-1/2}^{(1, 0)}(\eta) }{1+r}
    \end{pmatrix}\,,
\nn\\
  {\mathbf{h}}^{2}_{\oo,\ell}(r, \eta) &= 
   \frac{ v^{\frac{1}4(\Delta_{12}+\Delta_{43})} }
             {  \sqrt{1-r^2} \sqrt{1 + r^2 + 2 r \eta}} 
     \begin{pmatrix}
          \frac{ P_{\ell-1/2}^{(0, 1)}(\eta)}{1-r}  -\frac{P_{\ell-1/2}^{(1, 0)}(\eta)}{1+r}  \\[5pt] 
         \frac{P_{\ell-1/2}^{(0, 1)}(\eta)}{1-r}   +  \frac{ P_{\ell-1/2}^{(1, 0)}(\eta) }{1+r}
    \end{pmatrix}\,,
   \nn \\
    {\mathbf{h}}^{3}_{\oo,\ell}(r, \eta) &= \frac{ v^{\frac{1}4(\Delta_{12}+\Delta_{43})} }{  \sqrt{1-r^2} \sqrt{1 + r^2 - 2 r \eta}} 
	\begin{pmatrix}    
       \frac{P_{\ell-1/2}^{(0, 1)}(\eta)}{1+r}  - \frac{P_{\ell-1/2}^{(1, 0)}(\eta)}{1-r}    \\[5pt] 
           \frac{ P_{\ell-1/2}^{(0, 1)}(\eta)}{1+r} + \frac{ P_{\ell-1/2}^{(1, 0)}(\eta) }{1-r}      
    \end{pmatrix} \,, 
   \nn \\
  {\mathbf{h}}^{4}_{\oo,\ell}(r, \eta) &= 
   \frac{ v^{\frac{1}4(\Delta_{12}+\Delta_{43})} }
             {  \sqrt{1-r^2} \sqrt{1 + r^2 - 2 r \eta}} 
     \begin{pmatrix}
         \frac{  P_{\ell-1/2}^{(0, 1)}(\eta)}{1+r}  +\frac{P_{\ell-1/2}^{(1, 0)}(\eta)}{1-r} 
                   \\[5pt]  
                 \frac{ P_{\ell-1/2}^{(0, 1)}(\eta)}{1+r}   -  \frac{ P_{\ell-1/2}^{(1, 0)}(\eta) }{1-r}
    \end{pmatrix} \,,
 \end{align}
 where $P^{(a,b)}_n(\eta)$ are Jacobi polynomials.

\subsection{Poles and Residues}

The Casimir equations \eqref{CasEqs} can be solved order by order in a series expansion in $r$ and $\eta$. The full solution presented below was guessed based on the first few terms in this expansion. It was then verified that this guess solves the Casimir equations to a very high order in the expansion in $r$ and $\eta$. 

In practice, the procedure for solving \eqref{CasEqs} in a series expansion and extracting the poles and residues appearing in \eqref{eq:recurssionFormula} is as follows. A solution of \eqref{CasEqs} for the functions $\mathbf{h}^{I_{\pm}}_{\Delta,\ell}=r^{-\Delta}\mathbf{g}^{I_{\pm}}_{\Delta,\ell}$ has an expansion
\begin{align}
\mathbf{h}^{I_{\pm}}_{\Delta,\ell}(r,\eta) = \sum_{k=0}^{\infty}\sum_{j=0}^{k+\ell-\frac{1}{2}} \mathbf{a}^{I_{\pm}}_{k,j}(\Delta,\ell)r^k\eta^j \ec
\end{align}
where the dependence on the external operator dimensions $\Delta_{12}$ and $\Delta_{34}$ is suppressed. The zeroth order coefficients $\mathbf{a}^{I_{\pm}}_{0,0}(\Delta)$ are fixed by the explicit solution for $\mathbf{h}^{I_{\pm}}_{\infty,\ell}(r,\eta)$ given in \eqref{h1Def}. For a given $\ell$ we solve for the coefficients $\mathbf{a}^{I_{\pm}}_{k,j}$ up to a finite order $k<\Lambda$, and rewrite the result in the form of the recursion formula \eqref{eq:recurssionFormula}. In particular, the locations of the poles $\Delta_A^{\star}$ are simply read off from the coefficients $\mathbf{a}^{I_{\pm}}_{k,j}(\Delta,\ell)$, while the spins $\ell_A$ are inferred from the degree of the polynomials in $\eta$ that multiply each pole.

Following the above procedure, we found three series of poles (two infinite and one finite), in analogy with the case of scalar blocks \cite{Kos:2013tga,Kos:2014bka,penedones2015recursion}.  They are listed in Table \ref{tableWithFermiPoles}.  We now describe the differential operators and residues associated with each pole.\footnote{Because we obtained the residues from the Casimir equation, and not by explicitly constructing null descendants, we cannot immediately fix the normalization of the differential operators $\cD_A$, and the corresponding values of $Q_A,M_A$.  Different normalization conventions will rescale $Q_A,M_A$ in such a way that the residue is unaffected. Our convention is such that $Q_A,M_A$ are as simple as possible.}

\begin{table}
\begin{center}
  \begin{tabular}{ |c|c | c | c|c|l|}
    \hline
   $A$ & $n_A$ & $\Delta_A^\star$ & $\ell_A$ & $\mathrm{parity}(\cD_A)$ & \textrm{values of $k$} \\ \hline\hline
    $(1,k)$ & $k$ & $1 - \ell- k$ & $\ell + k$ & $+$ & $ k = 1, 2, \ldots$\,,\\ 
    $(2,k)$ & $k$ & $(3-k)/2 $ & $\ell$ & $-$ & $k = 1,\, 3, \,5, \,\ldots$\,,\\ 
    $(3,k)$ & $k$ &  $2 + \ell   -  k$  & $\ell - k$ & $+$ & $k = 1, 2, \ldots, \ell - \frac{1}{2}$\,.\\\hline
  \end{tabular}
\end{center}
\caption{\label{tableWithFermiPoles}Table of poles in $\Delta$ for the fermion-scalar conformal blocks. There are three series of poles, which we call $A=(1,k), (2,k), (3,k)$. The integer $k$ ranges over the values shown. $n_A$ is the level of the descendant corresponding to the pole, $\Delta_A^\star$ gives the dimension at which the pole appears, and $\ell_A$ is the spin of the descendant associated with the pole. $\mathrm{parity}(\cD_A)$ indicates whether the differential operator $\cD_A$ preserves or changes parity.}
\end{table}

\begin{itemize}
\item \textit{First series of poles:} The differential operators with label $(1,k)$ have the form
\be
\cD_{1,k} &\propto (s \ptl_x s)^k \,,
\ee
where $s\ptl_x s=s_\alpha (\ptl_x)^\alpha{}_\beta s^\beta$.\footnote{The labels $(1,k),(2,k),(3,k)$ correspond to descendants of type I, III, II in \cite{penedones2015recursion}.}  By solving the Casimir equation, we find
\begin{gather}
Q_{1,k} =  - \frac{(-4)^{k}k\left(\frac 32 + \ell \right)_{k}}{(k!)^2(1+\ell)_{k}} \,,
\label{prefactor1}\nn\\
(M_{1,k}^{\De})_+^+ = \left(\frac{-2k - 2\Delta  + 1}4\right)_{k} \,, \qquad
(M_{1,k}^\De)_-^- = \left(\frac{-2 k - 2\Delta + 3}4\right)_{k}
,\qquad \nn\\
(M_{1,k}^\De)^+_- = (M_{1,k}^\De)^-_+ = 0 \,,
\end{gather}
where the Pochhammer symbol is defined by $(a)_{n} = \Gamma(a+n)/\Gamma(a)$.  The coefficients $(M_{1,k}^\De)^+_-, (M_{1,k}^\De)_-^+$ vanish because $\cD_{1,k}$ is parity even. The residue matrices $ \mathbf c_1^\pm$ are obtained by plugging the above into (\ref{eq:cmatrices}).  They are diagonal.

\item \textit{Second series of poles:} The differential operators with label $(2,k)$ have the form
\be
\cD_{2,k} &\propto \left( s\ptl_x\ptl_s \right) \sum_m a_m \ptl_x^{2m}((s \ptl_x s)(\ptl_s \ptl_x \ptl_s))^{n-m} \,, 
\ee
where the coefficients $a_m$ could be determined by demanding that $\cD_{2,k}\cO_{\De,\ell}$ is primary when $\De=\De_{2,k}^\star$. We find
\be
Q_{2,k} &= \frac{4^k k\Gamma(\frac{k}2)^2}{2\pi \Gamma\left(\frac{1+k}2 \right)^2 \left(\frac{1}2 +\ell - \frac{k}{2}\right)_{k} \left(\frac{3}2 +\ell - \frac{k}{2}\right)_{k}} \,, 
\nn\\
(M^\De_{2,k})^+_- &= \left(\frac{-k+2\ell-2\Delta+2}{4}\right)_{\frac{k+1}{2}} \left(\frac{-k+2\ell+2\Delta+4}{4}\right)_{\frac{k-1}{2}}\,, 
\nn\\
(M^\De_{2,k})^-_+ &= \left(\frac{-k-2\ell-2\Delta}{4}\right)_{\frac{k+1}{2}} \left(\frac{-k-2\ell+2\Delta+2}{4}\right)_{\frac{k-1}{2}}\,, \nn\\
(M_{2,k}^\De)^+_+ &= (M_{2,k}^\De)_-^- = 0 \,,
\ee
where the matrix $\mathbf c_2^\pm$ is anti-diagonal for all values of $k$ and $\ell$.

\item \textit{Third series of poles:} The differential operators with label $(3,k)$ have the form
\be
\cD_{3,k} &\propto (\ptl_s\ptl_x\ptl_s)^k \,.
\ee
We find
\begin{gather}
Q_{3,k} = -\frac{(-4)^k k\left(\frac 12  - k + \ell  \right)_k}{(k!)^2(1-k+\ell)_{k}} \,, \nn\\
(M_{3,k}^{\De})_+^+ = \left(\frac{-2 k - 2\Delta + 3}4\right)_{k}\,, \qquad
(M_{3,k}^\De)_-^- = \left(\frac{-2k - 2\Delta + 1}4\right)_{k}\,, \nn\\
(M_{3,k}^\De)^+_- = (M_{3,k}^\De)^-_+ = 0 \,,
\end{gather}
where the matrix $\mathbf{c}_3^\pm$ is thus diagonal for all values of $k$ and $\ell$.

\end{itemize}

\section{Summary and Discussion}
\label{DISCUSSION}

Our main result in this work is an explicit recursion relation for the conformal blocks corresponding to the exchange of a half-odd integer spin conformal primary that can be used to compute the blocks numerically.  This recursion formula takes the form \eqref{eq:recurssionFormula}, with the various quantities appearing in this formula determined in Section~\ref{results}.

As a check on our results, let us decompose a simple free theory four-point function into conformal blocks. Consider the theory of a free Majorana fermion $\psi$ and free scalar $\phi$ in three space-time dimensions.  We have
\be
\<\Psi(X_1,S_1)\Phi(X_2)\Phi(X_3)\Psi(X_4,S_4)\> &= \frac{i \<S_1 S_4\>}{X_{14}^{3/2} X_{23}^{1/2}} = \frac{t_1}{X_{14}X_{23}^{1/2}} \,,
\ee
so that
\be
g_\mathrm{free}^1(u,v) &= \frac{u^{3/4}}{v^{1/2}},\qquad g_\mathrm{free}^2(u,v) = 0 \,.
\ee
By matching several orders in an expansion in $r$, we find
\be
g^{I_+}_\mathrm{free}(u,v)
&= \sum_{s=0}^\oo p_s
g^{I_+,++}_{\frac 3 2 + s,\frac 1 2+s}(u,v) \,,
\qquad (I_+=1,2)\,,\\
p_s &= -4(s+1)\,.
\ee
Thus, the free four-point function can be expanded in conformal blocks for higher spin currents
\be
J_s^{\mu_1\cdots\mu_s} &= \psi \ptl^{\mu_1}\cdots\ptl^{\mu_s} \phi\,, \qquad (s\geq 0) \,.
\ee
Furthermore, only the parity even OPE coefficients $\lambda^+_{\Psi\Phi J_s}$ are nonzero. Each coefficient is pure imaginary since $p_s = (\lambda^+_{\Psi\Phi J_s})^2$ is negative, as expected in a unitary theory.

With the conformal blocks for mixed correlators between scalars and fermions in hand, one can now investigate the constraints that crossing symmetry and unitarity impose on the space of CFTs with both spin-$0$ and spin-$1/2$ conformal primary operators.  We hope to report on such a study in a future publication.

\section*{Acknowledgements}
We thank Jo\~ao Penedones, Emilio Trevisani, and Masahito Yamazaki for discussions. This work was supported by the US NSF under grant No.~PHY-1418069 (LI, SSP, and RY), DOE grant number DE-SC0009988 (DSD), and  NSF grant PHY-1350180 (FK and DP). DSD is supported in part by a William D. Loughlin Membership at the Institute for Advanced Study. DP is supported in part by a Martin A. and Helen Chooljian Founders' Circle Membership at the Institute for Advanced Study.

\appendix

\section{The Casimir Equation}
\label{AppendixCasEq}

In this Appendix we derive the Casimir equations satisfied by the conformal blocks and solve them at large $\Delta$.

For a scalar field $\Phi(x)$ whose embedding space parent (defined on the 5D light cone) is $\Phi(X) = \frac{1}{(X^+)^{\Delta_\Phi}} \Phi(x)$, we can write the action of the 5D Lorentz generators as
 \es{LorentzScalar}{
  i[J^{AB},\Phi(X)] = \left( X^B \frac{\partial}{\partial X_A} -  X^A\frac{\partial}{\partial X_B}   \right) \Phi(X) \,.
 }
Upon reduction to 3D, \eqref{LorentzScalar} reduces to the action of the conformal generators on the conformal primary field $\Phi(x)$ of dimension ${\Delta_\Phi}$.

For a 3D spinor field $\Psi^\alpha(x)$ whose embedding space parent is $\Psi^I(X)$, as defined in the main text through $S_I \Psi^I(X) =\frac{1}{(X^+)^{\Delta_\Psi}} s_\alpha \Psi^\alpha(x)$, the action of the conformal generators is more subtle.  The subtlety comes from the fact that, because the 5D spinor polarizations $S_I$ are transverse, $S_I X^I{}_J = 0$, the parent spinor field $\Psi^I(X)$ is defined only modulo shifts of the form 
 \es{PsiAmbiguity}{
   \Psi^I(X) \to \Psi^I(X) + X^I{}_J \Theta^J(X) \,, 
 }
where $\Theta^J(X)$ is an arbitrary spinor on the 5D light cone.  To remove this ambiguity, one can define 
 \es{TildePsiDef}{
   \tilde \Psi^I(X) = X^I{}_J \Psi^I(X) \,.
 }
The 5D $\tilde \Psi^I(X)$ does not suffer from the ambiguity in \eqref{PsiAmbiguity} because $X^I{}_J X^J{}_K = 0$ on the 5D light cone.  On $\tilde \Psi$, the 5D Lorentz generators have the usual action in the spinor representation
 \es{tildePsiAction}{
i[J^{AB},\tilde \Psi(X)] = \left( X^B\frac{\partial}{\partial X_A} -  X^A\frac{\partial}{\partial X_B}  + \dfrac{1}{4}\left[\Gamma^A, \Gamma^B\right] \right) \tilde \Psi(X) \,,
 }
where we have suppressed all spinor indices.  One can check that, once traced through all the definitions, Eq.~\eqref{tildePsiAction} implies that the 3D spinor field $\Psi^\alpha(x)$ is a conformal primary of dimension $\Delta_\Psi$.

Let us now analyze the conformal block decomposition of the mixed 4-point function $\langle \Psi_1(X_1)  \Phi_2(X_2)  \Phi_3(X_3)  \Psi_4(X_4) \rangle$ in \eqref{mixedGeneralForm}.  Using \eqref{TildePsiDef} and \eqref{mixedGeneralForm}--\eqref{eq:parityOddBasis}, we can write
 \es{FourPtTilde}{
  \langle \tilde \Psi_1^J(X_1)  \Phi_2(X_2)  \Phi_3(X_3)  \tilde \Psi_4^K(X_4) \rangle
   = \left(\dfrac{X_{24}}{X_{14}} \right)^{
     \frac{\Delta_{12}}2} \left(\dfrac{X_{13}}{X_{14}} \right)^{
     \frac{\Delta_{43}}2}
     \frac{\sum_I (\tilde t_I)^{JK} \,  g^I(u,v)}{X_{12}^{\frac{\Delta_1+\Delta_2}2} 
     X_{34}^{\frac{\Delta_3+\Delta_4}2}} \,, 
 }
where 
 \es{tTildeDef}{
   (\tilde t_1)^{JK} &=  -i \frac{(X_1 X_4)^{JK}}{\sqrt{X_{14}}}\,,\qquad
   (\tilde t_2)^{JK}=   i \frac{(X_1 X_2 X_3 X_4)^{JK}}{\sqrt{X_{12} X_{34} X_{23}}} \,, \\
   (\tilde t_3)^{JK} &=  i \frac{(X_1 X_2 X_4)^{JK}}{\sqrt{X_{12} X_{24} }} \,, \qquad
   (\tilde t_4)^{JK} =  i \frac{(X_1 X_3 X_4)^{JK} }{\sqrt{X_{13} X_{34}}} \,.
 }
Let $(J_1)^{KL}$ and $(J_2)^{KL}$ be the 5D Lorentz generators acting on $\tilde \Psi_1$ and $\Phi_2$ according to \eqref{tildePsiAction} and \eqref{LorentzScalar}, respectively, and let us denote their sum squared as
\begin{equation}
 L^2 = \dfrac{1}{2}(J_1 + J_2)^{KL} (J_1 + J_2)_{KL} \,.
\end{equation}
It can be shown that the contribution to \eqref{FourPtTilde} coming from the conformal multiplet of a primary ${\cal O}_{\Delta, \ell}$ must obey the eigenvalue equation 
 \es{EvalueEq}{
 (L^2 -  C_{\Delta, \ell})\left(\left(\dfrac{X_{24}}{X_{14}} \right)^{
\frac{\Delta_{12}}2} \left(\dfrac{X_{13}}{X_{14}} \right)^{
\frac{\Delta_{43}}2}  \sum_I \frac{\tilde t_I g^I_{\Delta, \ell}(u, v)}{X_{12}^{\frac{\Delta_1+\Delta_2}2 }X_{34}^{\frac{\Delta_3+\Delta_4}2}} \right) = 0 \,,
 }
with eigenvalue
 \es{Evalue}{
  C_{\Delta, \ell} =  \Delta(\Delta - 3) + \ell(\ell+1) \,.
 }

Using the explicit expressions in \eqref{tildePsiAction} and \eqref{LorentzScalar}, one can write down the equations for $g^{I}_{\Delta, \ell}$ implied by \eqref{EvalueEq} as
 \es{CasEqs}{
  \left[  
   \begin{pmatrix}
   {\cal L}_D^+ & {\cal L}_A^+ \\
   {\cal L}_A^+ & {\cal L}_D^+ 
  \end{pmatrix} 
  + \begin{pmatrix}
    0 & \frac{4r(\Delta_{12} + \Delta_{43})}{1 + r^2 - 2 r \eta}\\
    0 & \frac{4r(\eta - 2 r + r^2 \eta) (\Delta_{12} + \Delta_{43})}{(1 + r^2 - 2 r \eta)^2}
  \end{pmatrix} \right] 
  \begin{pmatrix}
   g^1_{\Delta, \ell} \\
   g^2_{\Delta, \ell}
  \end{pmatrix} &= C_{\Delta, \ell} \begin{pmatrix}
   g^1_{\Delta, \ell} \\
   g^2_{\Delta, \ell}
  \end{pmatrix}, \\
   \left[  
   \begin{pmatrix}
   {\cal L}_D^- & {\cal L}_A^- \\
   {\cal L}_A^- & {\cal L}_D^- 
  \end{pmatrix} 
  + \begin{pmatrix}
    \frac{4r(\eta + 2 r + r^2 \eta)  \Delta_{43}}{(1 + r^2 + 2 r \eta)^2} & -\frac{4r \Delta_{12} }{1 + r^2 + 2 r \eta}\\
     -\frac{4r \Delta_{43} }{1 + r^2 + 2 r \eta} & \frac{4r(\eta + 2 r + r^2 \eta)  \Delta_{12}}{(1 + r^2 + 2 r \eta)^2}
  \end{pmatrix} \right] 
  \begin{pmatrix}
   g^3_{\Delta, \ell} \\
   g^4_{\Delta, \ell}
  \end{pmatrix} &= C_{\Delta, \ell} \begin{pmatrix}
   g^3_{\Delta, \ell} \\
   g^4_{\Delta, \ell}
  \end{pmatrix},
 }
where
 \es{DiffOpsExplicit}{
  {\cal L}_D^\pm &= r^2 \partial_r^2 + (\eta^2 - 1) \partial_\eta^2 \\
   &{}+ \Biggl( \frac{4r^2\eta(1-r^2)(\Delta_{12} +\Delta_{43})}{(1 + r^2 -2r\eta)(1 + r^2 +2r\eta)}  - \frac{r(1+3r^2)}{1-r^2}
   - \frac{r ( 1 - r^2)(1 + r^2 \mp 2 r \eta)}{(1 + r^2 + 2 r \eta)(1 + r^2 - 2 r \eta)} \Biggr) \partial_r \\
   &{}+ \left( \frac{4 \left(\eta ^2-1\right) \left(r^3+r\right) (\Delta_{12} + \Delta_{43})}{\left(1+r^2+2 \eta  r\right) \left(1+r^2-2 \eta  r\right)} 
    + \frac{\left[ 3 \eta (1 + r^2) \pm 2 r(4 \eta^2 - 1) \right] (1 + r^2 \mp 2 r \eta)}{(1 + r^2 + 2 r \eta) (1 + r^2 - 2 r \eta)}\right) \partial_\eta \\
   &{}+ \left( \frac 34 - \frac{4r \Delta_{12} \Delta_{43} \left(\eta + 2 r + r^2 \eta \right)}{(1 + r^2 + 2 r \eta)^2} \right), \\
  {\cal L}_A^\pm &= \frac{2r^2}{1 - r^2} \partial_r \pm \partial_\eta \,,
 }
and the coordinates $r$ and $\eta$ are defined in \eqref{uvToreta}.  Note that the parity-even blocks, $g^1_{\Delta, \ell}$ and $g^2_{\Delta, \ell}$, decouple from the parity-odd blocks, $g^3_{\Delta, \ell}$ and $g^4_{\Delta, \ell}$.

At large $\Delta$, these equations can be solved as follows.  From \eqref{Recursion}, we expect 
 \es{gToh}{
  g_{\Delta, \ell}^I (r, \eta) = r^\Delta \left[ h_{\infty, \ell}^I(r, \eta) + O(1/\Delta, r) \right] \,,
 }
for some functions $h_{\infty, \ell}^I$ independent of $\Delta$ that we now find.  Note that the term denoted by $O(1/\Delta, r)$ in \eqref{gToh} decays not only at large $\Delta$ but also at small $r$.  (Recall that $n_A \geq 1$ in \eqref{Recursion}.)  Plugging \eqref{gToh} into \eqref{CasEqs} and expanding up to order $\Delta^1$ at large $\Delta$, we find 
 \es{CasEqsApprox}{ 
   \begin{pmatrix}
   {\cal D}^+ & \frac{2r^2}{1 - r^2} \\
   \frac{2r^2}{1 - r^2} & {\cal D}^+ 
  \end{pmatrix}  
  \begin{pmatrix}
   h^1_{\infty, \ell} \\
   h^2_{\infty, \ell}
  \end{pmatrix} &= \begin{pmatrix}
   0 \\
   0
  \end{pmatrix}\,, \\
   \begin{pmatrix}
   {\cal D}^- & \frac{2r^2}{1 - r^2} \\
   \frac{2r^2}{1 - r^2} & {\cal D}^- 
  \end{pmatrix} 
  \begin{pmatrix}
   h^3_{\infty, \ell} \\
   h^4_{\infty, \ell}
  \end{pmatrix} &= \begin{pmatrix}
   0 \\
   0
  \end{pmatrix} \,,
 }
with
 \es{DiffOpsExplicitApprox}{
  {\cal D}^\pm &= 2 r \partial_r +2  
   + \frac{4r^2\eta(1-r^2)(\Delta_{12} +\Delta_{43})}{(1 + r^2 -2r\eta)(1 + r^2 +2r\eta)}  - \frac{r(1+3r^2)}{1-r^2} \\
   &{}- \frac{r ( 1 - r^2)(1 + r^2 \mp 2 r \eta)}{(1 + r^2 + 2 r \eta)(1 + r^2 - 2 r \eta)}   \,.
 }
These are first order ordinary differential equations in $r$.   Their solution is 
 \es{hFirst}{
  h_{\infty, \ell}^1(r, \eta) &=  \frac{ v^{\frac{1}4(\Delta_{12}+\Delta_{43})} }
             { \sqrt{1-r^2} \sqrt{1 + r^2 + 2 r \eta}} \left( \frac{ -c_1(\eta)}{1-r}  -\frac{c_2(\eta)}{1+r}  \right) \,, \\
  h_{\infty, \ell}^2(r, \eta) &=  \frac{ v^{\frac{1}4(\Delta_{12}+\Delta_{43})} }
             { \sqrt{1-r^2} \sqrt{1 + r^2 + 2 r \eta}} \left( \frac{ c_1(\eta)}{1-r}  -\frac{c_2(\eta)}{1+r}  \right) \,, \\    
  h_{\infty, \ell}^3(r, \eta) &=  \frac{ v^{\frac{1}4(\Delta_{12}+\Delta_{43})} }
             { \sqrt{1-r^2} \sqrt{1 + r^2 - 2 r \eta}} \left( \frac{ c_3(\eta)}{1+r}  -\frac{c_4(\eta)}{1-r}  \right) \,, \\ 
  h_{\infty, \ell}^4(r, \eta) &=  \frac{ v^{\frac{1}4(\Delta_{12}+\Delta_{43})} }
             { \sqrt{1-r^2} \sqrt{1 + r^2 - 2 r \eta}} \left( \frac{ c_3(\eta)}{1+r}  + \frac{c_4(\eta)}{1-r}  \right) \,,
 }
where $c_i(\eta)$ are integration constants that are arbitrary functions of $\eta$.

To determine $c_i(\eta)$, one has to expand the Casimir equations \eqref{CasEqs} to next-to-leading order in $1/\Delta$, and the result to order $r^0$.  Such an expansion makes sense because the terms denoted by $O(\Delta^0, r)$  in \eqref{gToh}  do not contribute in this limit.  We find that the $c_i$'s must obey the second order differential equations 
 \es{cDiff}{
  \left[(1-\eta^2) \partial_\eta^2 + (1 - 3 \eta) \partial_\eta  + \left( \ell + \frac 32 \right) \left( \ell -\frac 12 \right) \right]
   \begin{pmatrix}
    c_1(\eta) \\
    c_3(\eta) 
   \end{pmatrix} &= \begin{pmatrix} 0 \\ 0 \end{pmatrix} \,, \\
   \left[(1-\eta^2) \partial_\eta^2 + (1 + 3 \eta) \partial_\eta  + \left( \ell + \frac 32 \right) \left( \ell -\frac 12 \right) \right]
   \begin{pmatrix}
    c_2(\eta) \\
    c_4(\eta) 
   \end{pmatrix} &= \begin{pmatrix} 0 \\ 0 \end{pmatrix} \,.
 }
Each differential equation in \eqref{cDiff} has only one normalizable solution:
 \es{cSoln}{
  c_1(\eta) &= C_1 P_{\ell-1/2}^{(0, 1)}(\eta) \,, \qquad
   c_2(\eta) = C_2 P_{\ell-1/2}^{(1, 0)}(\eta) \,, \\
  c_3(\eta) &= C_3 P_{\ell-1/2}^{(0, 1)}(\eta) \,, \qquad
   c_4(\eta) = C_4 P_{\ell-1/2}^{(1, 0)}(\eta) \,,
 }
with $C_i$ being arbitrary constants and $P_n^{(a, b)}(\eta)$ being Jacobi polynomials.  In \eqref{h1Def}, we choose $C_1 = C_3 = 1$ and $C_2 = C_4 = 1$ (top entry of each two-component vector) or $C_2 = C_4 = -1$ (bottom entry of each two-component vector).

\section{Determining OPE Coefficients}
\label{AppendixRelation3Pt4Pt}

The  OPE coefficients $\lambda^\pm_{\Psi \Phi \cO}$ appearing in the three-point function (\ref{eq:threePointFunc}) are the same as those appearing in the four-point function \eqref{eq:blockgI} only for a specific normalization of the operator ${\cal O}$.  If we write the two point function of $\cO$ as
\be
\label{eq:twoPointFunc}
\<\cO^{\alpha_1 \ldots \alpha_{2l}}(x_1) \cO^{\beta_1 \ldots \beta_{2l}}(x_2) \> = i^{2\ell} c_{\cO}\frac{(x_{12}i \sigma_2)^{\alpha_1(\beta_1} \cdots (x_{12}i \sigma_2)^{|\alpha_{2\ell}|\beta_{2\ell})} }{|x_{12}|^{2\Delta+2\ell}} \,,
\ee
where $x_{ij}=x_i - x_j$ and $\alpha_1, \ldots ,\alpha_{2\ell}$ and $\beta_1, \ldots ,\beta_{2\ell}$ are spinor indices, our task is therefore to determine $c_{\cO}$.\footnote{In a reflection positive theory, the OPE coefficients $\lambda_{\Psi_i \Phi_j \cO}$ will be pure imaginary and $c_\cO$ will be positive.  In a theory that violates reflection positivity, our conventions are such that $c_\cO$ should still be fixed to the positive value determined in this appendix, after which the coefficients $\lambda_{\Psi_i \Phi_j \cO}$ may no longer be pure imaginary.}

Recall that the 3-point function $\Psi_1 \Phi_2 {\cal O}$ is given in \eqref{eq:threeptstructures}. 
Explicitly, we have
 \es{eq:threePointFuncRealSpace}{
 &\<\Psi_1^\beta(x_1) \Phi_2(x_2)\cO^{\alpha_1 \alpha_2 \ldots \alpha_{2\ell}}(x_3)\> \\
 &= \lambda^+_{\Psi_1 \Phi_2 \cO} \frac{(x_{13}i \sigma_2)^{\beta (\alpha_1} (x_{31}x_{12}x_{23}i \sigma_2)^{\alpha_2\alpha_3}\cdots (x_{31}x_{12}x_{23}i \sigma_2)^{\alpha_{2\ell-1}\alpha_{2\ell})}}{|x_{12}|^{\Delta_{1} + \Delta_2 - \Delta + \ell-\frac{1}2 }|x_{23}|^{\Delta_2 + \Delta - \Delta_{1}+\ell - \frac{1}2}|x_{31}|^{\Delta + \Delta_{1} - \Delta_2+\ell + \frac{1}2}} \\ 
  &{}- \lambda^-_{\Psi_1 \Phi_2 \cO} \frac{(x_{12}x_{23}i \sigma_2)^{\beta (\alpha_1} (x_{31}x_{12}x_{23}i \sigma_2)^{\alpha_2\alpha_3}\cdots(x_{31}x_{12}x_{23}i \sigma_2)^{\alpha_{2l-1}\alpha_{2l})}}{|x_{12}|^{\Delta_{1} + \Delta_2 - \Delta +\ell+\frac{1}2}|x_{23}|^{\Delta_2 + \Delta - \Delta_{1}+\ell +\frac{1}2}|x_{31}|^{\Delta + \Delta_{1} - \Delta_2+\ell-\frac{1}2 }} \,.
}
In the OPE limit $x_1 \rightarrow x_2$, the three-point function \eqref{eq:threePointFuncRealSpace} becomes
 \es{eq:mixedOPElimit}{
  &\<\Psi_1^\beta(x_1) \Phi_2(x_2)\cO^{\alpha_1 \alpha_2 \ldots \alpha_{2\ell}}(x_3)\> \\
  &\approx \lambda^+_{\Psi_1 \Phi_2 \cO} (i \sigma_2 x_{12})_{\beta_1 \beta_2} \cdots (i \sigma_2 x_{12})_{\beta_{2\ell-2} \beta_{2\ell-1}}\frac{(x_{23}i \sigma_2)^{\beta (\alpha_1} (x_{23}i \sigma_2)^{|\beta_1|\alpha_2} \cdots (x_{23}i \sigma_2)^{|\beta_{2\ell-1}|\alpha_{2\ell})}}{|x_{12}|^{\Delta_{1} + \Delta_2 - \Delta +\ell-\frac{1}2}|x_{23}|^{ 2\Delta +2\ell }} \\ 
  &{}- \lambda^-_{\Psi_1 \Phi_2 \cO} (i \sigma_2 x_{12})_{\beta_1 \beta_2} \cdots (i \sigma_2 x_{12})_{\beta_{2\ell-2} \beta_{2\ell-1}}\frac{(x_{12}x_{23}i \sigma_2)^{\beta (\alpha_1} (x_{23}i \sigma_2)^{|\beta_1|\alpha_2} \cdots (x_{23}i \sigma_2)^{|\beta_{2\ell-1}|\alpha_{2\ell})}}{|x_{12}|^{\Delta_{1} + \Delta_2 - \Delta+ \ell+\frac{1}2}|x_{23}|^{ 2\Delta +2\ell}} \,.
}
From \eqref{eq:mixedOPElimit}, we can extract the contribution of ${\cal O}$ to the $\Psi_1 \times \Phi_2$ OPE:
 \es{GotOPE}{
  & \Psi^\beta_1(x_1)  \Phi_2(x_2)  \\ 
  &= \cdots +  \bigg(  -\frac{i^{2\ell}\lambda^+_{\Psi_1 \Phi_2 \cO} }{c_\cO}(i\sigma_2 x_{12})_{\alpha_1\beta_1} \cdots (i\sigma_2 x_{12})_{\alpha_{\ell-\frac{1}2}\beta_{\ell-\frac{1}2}} \frac{\cO^{\beta \alpha_1 \beta_1 \ldots \alpha_{\ell-\frac{1}2} \beta_{\ell-\frac{1}2}}(x_2)}{|x_{12}|^{\Delta_{1} + \Delta_2 - \Delta +\ell-\frac{1}2}} \\ 
  &{}+  \frac{i^{2\ell}\lambda^-_{\Psi_1 \Phi_2 \cO} }{c_\cO}  (i\sigma_2 x_{12})_{\alpha_1\beta_1} \cdots (i\sigma_2 x_{12})_{\alpha_{\ell-\frac{1}2}\beta_{\ell-\frac{1}2}} (x_{12})^{\beta}_{\,\,\gamma}\frac{\cO^{\gamma \alpha_1 \beta_1 \ldots \alpha_{\ell-\frac{1}2} \beta_{\ell-\frac{1}2}}(x_2)}{ |x_{12}|^{\Delta_{1} + \Delta_2 - \Delta +\ell+\frac{1}2}}\bigg) + \cdots \,.
 }

Now we can examine the mixed 4-point function in the OPE limit.  As we take $x_3 \to x_4$, using \eqref{GotOPE} and \eqref{eq:twoPointFunc} and the fact that Eq. (\ref{eq:threePointFunc}) implies $\lambda^b_{  \Psi_4 \Phi_3 \cO} = (-1)^{\ell-\frac 12} \lambda^b_{  \Phi_3 \Psi_4 \cO} $, one can write the four-point function as
 \es{MixedFourPoint}{
  \< \Psi^\alpha_1(x_1)\Phi_2(x_2) \Phi_3 (x_3) \Psi^\beta_4(x_4)\> &=  
      \left(\dfrac{|x_{24}|}{|x_{14}|} \right)^{
      {\Delta_{12}}} \left(\dfrac{|x_{13}|}{|x_{14}|} \right)^{\Delta_{43}}      
        \sum_{ \cO}\sum_{a, b = \pm} 
       \frac{\lambda^a_{ \Psi_1  \Phi_2  \cO}  \lambda^b_{  \Psi_4 \Phi_3 \cO} \left(g^{ab}_\cO(x_i) \right)^{\alpha\beta}}{|x_{12}|^{\Delta_1+\Delta_2} |x_{34}|^{\Delta_3+\Delta_4}} \,,
   }
with
\be
\label{eq:g++}
 ( g^{++}_\cO)^{\alpha\beta}(x_i) &= \left(\dfrac{|x_{14}|}{|x_{24}|} \right)^{
{\Delta_{12}}} \left(\dfrac{|x_{14}|}{|x_{13}|} \right)^{
\Delta_{43}}  \frac{(-1)^{\ell + 1}}{c_\cO}(i\sigma_2 x_{34})_{\alpha_1\alpha_2} \cdots (i\sigma_2 x_{34})_{\alpha_{2\ell-2}\alpha_{2\ell-1}}  \nonumber \\
&\times \frac{(x_{23}i \sigma_2)^{\alpha (\beta} (x_{31}x_{12}x_{23}i \sigma_2)^{\alpha_1\alpha_2} \cdots (x_{31}x_{12}x_{23}i \sigma_2)^{\alpha_{2\ell-2}\alpha_{2\ell-1})}}{|x_{12}|^{- \Delta + \ell - \frac 12 }|x_{23}|^{ \Delta - \Delta_{12}+\ell - \frac{1}2}|x_{31}|^{\Delta+ \Delta_{12}+\ell + \frac{1}2} {|x_{34}|^{ - \Delta +\ell - \frac 12}} } \,,
\\
\label{eq:g--}
( g^{--}_\cO)^{\alpha\beta}(x_i) &=  \  \left(\dfrac{|x_{14}|}{|x_{24}|} \right)^{
{\Delta_{12}}} \left(\dfrac{|x_{14}|}{|x_{13}|} \right)^{
\Delta_{43}}  \frac{(-1)^{\ell }}{c_\cO}(i\sigma_2 x_{34})_{\alpha_1\alpha_2} \cdots (i\sigma_2 x_{34})_{\alpha_{2\ell-2}\alpha_{2\ell-1}}   \nonumber\\
&\times (x_{34})^\beta_{\,\,\gamma} \frac{(x_{12}x_{23}i \sigma_2)^{\alpha (\gamma} (x_{31}x_{12}x_{23}i \sigma_2)^{\alpha_1\alpha_2} \cdots (x_{31}x_{12}x_{23}i \sigma_2)^{\alpha_{2\ell-2}\alpha_{2\ell-1})}}{ |x_{12}|^{- \Delta + \ell + \frac 12}|x_{23}|^{\Delta - \Delta_{12}+\ell+ \frac{1}2}|x_{31}|^{\Delta + \Delta_{12}+\ell - \frac{1}2} {|x_{34}|^{- \Delta +\ell+\frac 12}} } \,,
\\
\label{eq:g+-}
( g^{+-}_\cO)^{\alpha\beta}(x_i) &= \left(\dfrac{|x_{14}|}{|x_{24}|} \right)^{
{\Delta_{12}}} \left(\dfrac{|x_{14}|}{|x_{13}|} \right)^{
\Delta_{43}}  \frac{(-1)^{\ell + 1}}{c_\cO}(i\sigma_2 x_{34})_{\alpha_1\alpha_2} \cdots (i\sigma_2 x_{34})_{\alpha_{2\ell-2}\alpha_{2\ell-1}} \nonumber  \\ &\times (x_{34})^\beta_{\,\,\gamma}  
 \frac{(x_{23}i \sigma_2)^{\alpha (\gamma} (x_{31}x_{12}x_{23}i \sigma_2)^{\alpha_1\alpha_2} \cdots (x_{31}x_{12}x_{23}i \sigma_2)^{\alpha_{2\ell-2}\alpha_{2\ell-1})}}{ |x_{12}|^{ - \Delta + \ell - \frac 12 }|x_{23}|^{\Delta - \Delta_{12}+\ell- \frac{1}2}|x_{31}|^{\Delta + \Delta_{12}+\ell+ \frac{1}2} {|x_{34}|^{ - \Delta +\ell+\frac 12}} } \,,
\\
\label{eq:g-+}
( g^{-+}_\cO)^{\alpha\beta}(x_i) &=   \left(\dfrac{|x_{14}|}{|x_{24}|} \right)^{
{\Delta_{12}}} \left(\dfrac{|x_{14}|}{|x_{13}|} \right)^{
\Delta_{43}}  \frac{(-1)^{\ell}}{c_\cO}(i\sigma_2 x_{34})_{\alpha_1\alpha_2}\cdots (i\sigma_2 x_{34})_{\alpha_{2\ell-2}\alpha_{2\ell-1}}  \nonumber \\
&\times \frac{(x_{12}x_{23}i \sigma_2)^{\alpha (\beta} (x_{31}x_{12}x_{23}i \sigma_2)^{\alpha_1\alpha_2} \cdots (x_{31}x_{12}x_{23}i \sigma_2)^{\alpha_{2\ell-2}\alpha_{2\ell-1})}}{ |x_{12}|^{- \Delta + \ell + \frac 12}|x_{23}|^{\Delta - \Delta_{12}+\ell+\frac{1}2}|x_{31}|^{\Delta + \Delta_{12}+\ell - \frac{1}2} {|x_{34}|^{- \Delta +\ell - \frac 12 }} } \,.
\ee

It is sufficient to write these expressions for a convenient choice of the $4$ coordinates $x_i$:
 \es{PointChoiceApp}{
x_1 &= (0, -1, 0)\,, \\
x_2 &= (0, 1, 0)\,, \\
x_3 &= (0, r \cos \theta, r \sin \theta)\,, \\
x_4 &= (0, - r \cos \theta, - r \sin \theta)\,.
 }
With this choice, the OPE limit $x_3\rightarrow x_4$ is realized as $r \rightarrow 0$.  In this limit, the structures that appear in \eqref{eq:g++}--\eqref{eq:g-+} can be approximated as
 \es{Limits}{
    i \sigma_2 x_{34} &\approx -2 r   (- \sigma_3 \cos \theta + \sigma_1 \sin \theta)\,, \qquad
      x_{23} i \sigma_2 \approx  -\sigma_3\,, \\
   x_{12}x_{23} i \sigma_2 &\approx 2 i \sigma_2\,, \text{\hspace{3.4cm}}
      x_{31}x_{12}x_{23}i \sigma_2 \approx 2 \sigma_3\,,
 }
while $|x_{12}|\approx 2$, $|x_{34}|\approx 2r$, $x_{14}\approx x_{23} \approx \sqrt{1+r^2 - 2r \cos \theta }$, and $x_{24}\approx x_{13} \approx \sqrt{1+r^2 + 2r \cos \theta}$.  Thus \eqref{eq:g++}--\eqref{eq:g-+} become
\be 
(g^{++}_\cO)^{\alpha\beta}(x_i)&\approx   (-1)^{\ell + 1} \frac{(4r)^{ \Delta}}{c_\cO}\nonumber (M_\theta)_{\alpha_1\alpha_2} \cdots (M_\theta)_{\alpha_{2l-2}\alpha_{2l-1}}
(-\sigma_3)^{\alpha(\beta} M_0^{\alpha_1\alpha_2} \cdots M_0^{\alpha_{2l-2}\alpha_{2l-1})}\,, \nonumber\\
( g^{--})^{\alpha\beta}_\cO(x_i)&\approx    (-1)^{\ell }\frac{(4r)^{ \Delta}}{ c_\cO} \nonumber (M_\theta)_{\alpha_1\alpha_2} \cdots (M_\theta)_{\alpha_{2l-2}\alpha_{2l-1}} (\tilde M_{\theta})^{\beta}_{\,\,\gamma} 
 (-i\sigma_2)^{\alpha(\gamma} M_0^{\alpha_1\alpha_2} \cdots M_0^{\alpha_{2l-2}\alpha_{2l-1})} \nonumber\,, \\
( g^{+-})^{\alpha\beta}_\cO(x_i)&\approx   (-1)^{\ell + 1} \frac{(4r)^{ \Delta}}{ c_\cO}\nonumber (M_\theta)_{\alpha_1\alpha_2} \cdots (M_\theta)_{\alpha_{2l-2}\alpha_{2l-1}} (\tilde M_{\theta})^{\beta}_{\,\,\gamma} 
 (-\sigma_3)^{\alpha(\gamma} M_0^{\alpha_1\alpha_2} \cdots M_0^{\alpha_{2l-2}\alpha_{2l-1})}\nonumber\,,\\
( g^{-+})^{\alpha\beta}_\cO(x_i)&\approx    (-1)^{\ell }\frac{(4r)^{ \Delta}}{ c_\cO}  (M_\theta)_{\alpha_1\alpha_2} \cdots (M_\theta)_{\alpha_{2l-2}\alpha_{2l-1}}  
 (-i \sigma_2)^{\alpha(\beta} M_0^{\alpha_1\alpha_2} \cdots M_0^{\alpha_{2l-2}\alpha_{2l-1})}\,,
\label{eq:gAssympLimit}
\ee
where $M_\theta = \sigma_3 \cos \theta - \sigma_1 \sin \theta$ and  $\tilde M_\theta = \sigma_3 \sin \theta+\sigma_1 \cos \theta$. 

This expression should be compared with our 4-point function in a similar limit.  Projected down to 3D, the 4-point function \eqref{mixedGeneralForm} takes the form: 
 \es{eq:mixedGeneralFormRealSpace}{
\<\Psi^\alpha_1(x_1)\Phi_2(x_2) \Phi_3 (x_3)\Psi^\beta_4(x_4)\> &= \left(\dfrac{|x_{24}|}{|x_{14}|} \right)^{
{\Delta_{12}}} \left(\dfrac{|x_{13}|}{|x_{14}|} \right)^{
\Delta_{43}}  
  \frac{\sum_I  t_I^{\alpha\beta} g^I(u, v)}{|x_{12}|^{\Delta_1+\Delta_2} |x_{34}|^{\Delta_3+\Delta_4}}\,,
  }
with the basis elements  being
 \es{eq:structureRealSpace}{
   t_1^{\alpha\beta} &= i \frac{(x_{14} i \sigma_2)^{\alpha \beta}}{|x_{14}|}, 
   \qquad  
   t_2^{\alpha\beta} = -  i \frac{(x_{12}x_{23}x_{34}i\sigma_2)^{\alpha\beta}}{|x_{12}||x_{23}||x_{34}|}\,, \\
   t_3^{\alpha\beta} &= i\frac{(x_{12}x_{24} i \sigma_2)^{\alpha\beta}}{|x_{12}||x_{24}|}\,, \qquad
   t_4^{\alpha\beta} = i  \frac{(x_{13}x_{34} i \sigma_2)^{\alpha\beta}}{|x_{13}||x_{34}|}\,.
 }
Considering the point choice \eqref{PointChoiceApp} and taking $r\to 0$, we have  
 \es{eq:structApproxOPELimit}{
   t_1 &\approx i \sigma_3, \qquad   t_2\approx i (- \sigma_3 \cos \theta + \sigma_1 \sin \theta)\,, \\   
   t_3 &\approx \sigma_2, \qquad   t_4\approx \sigma_2 \cos \theta- i \mathbf{1}_2 \sin \theta\,.
 }
In this limit, the functions appearing in the four-point function (\ref{eq:mixedGeneralFormRealSpace}) are simply given $g^I_{\Delta, \ell}(r, \eta) = r^\Delta h^{(\infty)I}_{\Delta, \ell}(r, \eta) + O(r^{\Delta+1})$. Using the first order expansion in $r$ for  $h^{I_+}_{\infty, \ell}(r, \eta)$ and for $h^{I_-}_{\infty, \ell}(r, \eta)$  (\ref{h1Def}) in the limit $r \rightarrow 0$, together with (\ref{eq:structApproxOPELimit}) and (\ref{eq:gAssympLimit}), we find that in the OPE limit 
\be
  \begin{pmatrix}
   ( g^{+ + })^{\alpha\beta}_{\cal O}(x_i) \\
   ( g^{-- })^{\alpha\beta}_{\cal O}(x_i)
   \end{pmatrix} &\approx \dfrac{4^{\Delta-1 }  (\ell+\frac 1 2)!}{(\frac 1 2)_{\ell+\frac 1 2}} \frac 1 {c_\cO} \sum_{I_+}t_{I_+}^{\alpha\beta} \mathbf g_{\Delta, \ell}^{I_+}(u, v)
    \,, \\
  \begin{pmatrix}
    ( g^{+ - })^{\alpha\beta}_{\cal O}(x_i)\\
    ( g^{-+ })^{\alpha\beta}_{\cal O}(x_i)
   \end{pmatrix} &\approx
   \dfrac{4^{\Delta-1 }  (\ell+\frac 1 2)!}{(\frac 1 2)_{\ell+\frac 1 2}} \frac 1 {c_\cO}  \sum_{I_-}t_{I_-}^{\alpha\beta} \mathbf g_{\Delta, \ell}^{I_-}(u, v) \,,
\ee
where $\mathbf{g}_{\Delta,\ell}^I\approx r^\De \mathbf{h}^I_{\oo,\ell}$ with $\mathbf{h}^I_{\oo,\ell}$  given by (\ref{h1Def}). 
In order for the OPE coefficients appearing in the three (\ref{eq:threePointFunc}) and four-point (\ref{eq:blockgI}) functions to be identical, we must choose
\be
c_\cO = \frac{4^{\Delta-1} (\ell+\frac 1 2)!}{\left(\frac 12 \right)_{\ell+\frac 12}} \,.
\label{eq:normalizationCond}
\ee

\bibliography{Biblio}{}

\begingroup\raggedright\begin{thebibliography}{10}

\bibitem{Ferrara:1973yt}
S.~Ferrara, A.~F. Grillo, and R.~Gatto, ``{Tensor representations of conformal
  algebra and conformally covariant operator product expansion},'' {\em Annals
  Phys.} {\bf 76} (1973) 161--188.

\bibitem{Polyakov:1974gs}
A.~M. Polyakov, ``{Nonhamiltonian approach to conformal quantum field
  theory},'' {\em Zh. Eksp. Teor. Fiz.} {\bf 66} (1974) 23--42.

\bibitem{Mack:1975jr}
G.~Mack, ``{Duality in Quantum Field Theory},'' {\em Nucl.Phys.} {\bf B118}
  (1977) 445.

\bibitem{Rattazzi:2008pe}
R.~Rattazzi, V.~S. Rychkov, E.~Tonni, and A.~Vichi, ``{Bounding scalar operator
  dimensions in 4D CFT},'' {\em JHEP} {\bf 12} (2008) 031,
  \href{http://xxx.lanl.gov/abs/0807.0004}{{\tt 0807.0004}}.

\bibitem{Rychkov:2009ij}
V.~S. Rychkov and A.~Vichi, ``{Universal Constraints on Conformal Operator
  Dimensions},'' {\em Phys. Rev.} {\bf D80} (2009) 045006,
  \href{http://xxx.lanl.gov/abs/0905.2211}{{\tt 0905.2211}}.

\bibitem{Caracciolo:2009bx}
F.~Caracciolo and V.~S. Rychkov, ``{Rigorous Limits on the Interaction Strength
  in Quantum Field Theory},'' {\em Phys. Rev.} {\bf D81} (2010) 085037,
  \href{http://xxx.lanl.gov/abs/0912.2726}{{\tt 0912.2726}}.

\bibitem{Poland:2010wg}
D.~Poland and D.~Simmons-Duffin, ``{Bounds on 4D Conformal and Superconformal
  Field Theories},'' {\em JHEP} {\bf 1105} (2011) 017,
  \href{http://xxx.lanl.gov/abs/1009.2087}{{\tt 1009.2087}}.

\bibitem{Rattazzi:2010gj}
R.~Rattazzi, S.~Rychkov, and A.~Vichi, ``{Central Charge Bounds in 4D Conformal
  Field Theory},'' {\em Phys. Rev.} {\bf D83} (2011) 046011,
  \href{http://xxx.lanl.gov/abs/1009.2725}{{\tt 1009.2725}}.

\bibitem{Rattazzi:2010yc}
R.~Rattazzi, S.~Rychkov, and A.~Vichi, ``{Bounds in 4D Conformal Field Theories
  with Global Symmetry},'' {\em J. Phys.} {\bf A44} (2011) 035402,
  \href{http://xxx.lanl.gov/abs/1009.5985}{{\tt 1009.5985}}.

\bibitem{Vichi:2011ux}
A.~Vichi, ``{Improved bounds for CFT's with global symmetries},'' {\em JHEP}
  {\bf 1201} (2012) 162, \href{http://xxx.lanl.gov/abs/1106.4037}{{\tt
  1106.4037}}.

\bibitem{Poland:2011ey}
D.~Poland, D.~Simmons-Duffin, and A.~Vichi, ``{Carving Out the Space of 4D
  CFTs},'' {\em JHEP} {\bf 1205} (2012) 110,
  \href{http://xxx.lanl.gov/abs/1109.5176}{{\tt 1109.5176}}.

\bibitem{Rychkov:2011et}
S.~Rychkov, ``{Conformal Bootstrap in Three Dimensions?},''
  \href{http://xxx.lanl.gov/abs/1111.2115}{{\tt 1111.2115}}.

\bibitem{ElShowk:2012ht}
S.~El-Showk, M.~F. Paulos, D.~Poland, S.~Rychkov, D.~Simmons-Duffin, and
  A.~Vichi, ``{Solving the 3D Ising Model with the Conformal Bootstrap},'' {\em
  Phys.Rev.} {\bf D86} (2012) 025022,
  \href{http://xxx.lanl.gov/abs/1203.6064}{{\tt 1203.6064}}.

\bibitem{Liendo:2012hy}
P.~Liendo, L.~Rastelli, and B.~C. van Rees, ``{The Bootstrap Program for
  Boundary CFT${}_d$},'' {\em JHEP} {\bf 1307} (2013) 113,
  \href{http://xxx.lanl.gov/abs/1210.4258}{{\tt 1210.4258}}.

\bibitem{Beem:2013qxa}
C.~Beem, L.~Rastelli, and B.~C. van Rees, ``{The $\mathcal{N}=4$ Superconformal
  Bootstrap},'' {\em Phys.Rev.Lett.} {\bf 111} (2013) 071601,
  \href{http://xxx.lanl.gov/abs/1304.1803}{{\tt 1304.1803}}.

\bibitem{Kos:2013tga}
F.~Kos, D.~Poland, and D.~Simmons-Duffin, ``{Bootstrapping the $O(N)$ vector
  models},'' {\em JHEP} {\bf 1406} (2014) 091,
  \href{http://xxx.lanl.gov/abs/1307.6856}{{\tt 1307.6856}}.

\bibitem{El-Showk:2013nia}
S.~El-Showk, M.~Paulos, D.~Poland, S.~Rychkov, D.~Simmons-Duffin, and A.~Vichi,
  ``Conformal Field Theories in Fractional Dimensions,'' {\em Phys. Rev. Lett.}
  {\bf 112} (Apr, 2014) 141601, \href{http://xxx.lanl.gov/abs/1309.5089}{{\tt
  1309.5089}}.

\bibitem{Alday:2013opa}
L.~F. Alday and A.~Bissi, ``{The superconformal bootstrap for structure
  constants},'' {\em JHEP} {\bf 09} (2014) 144,
  \href{http://xxx.lanl.gov/abs/1310.3757}{{\tt 1310.3757}}.

\bibitem{Gaiotto:2013nva}
D.~Gaiotto, D.~Mazac, and M.~F. Paulos, ``{Bootstrapping the 3d Ising twist
  defect},'' {\em JHEP} {\bf 1403} (2014) 100,
  \href{http://xxx.lanl.gov/abs/1310.5078}{{\tt 1310.5078}}.

\bibitem{Bashkirov:2013vya}
D.~Bashkirov, ``{Bootstrapping the $\mathcal{N}=1$ SCFT in three dimensions},''
  \href{http://xxx.lanl.gov/abs/1310.8255}{{\tt 1310.8255}}.

\bibitem{Berkooz:2014yda}
M.~Berkooz, R.~Yacoby, and A.~Zait, ``{Bounds on $\mathcal{N} = 1$
  superconformal theories with global symmetries},'' {\em JHEP} {\bf 1408}
  (2014) 008, \href{http://xxx.lanl.gov/abs/1402.6068}{{\tt 1402.6068}}.

\bibitem{El-Showk:2014dwa}
S.~El-Showk, M.~F. Paulos, D.~Poland, S.~Rychkov, D.~Simmons-Duffin, and
  A.~Vichi, ``{Solving the 3d Ising Model with the Conformal Bootstrap II.
  $c$-Minimization and Precise Critical Exponents},'' {\em J.Stat.Phys.} {\bf
  157} (June, 2014) 869, \href{http://xxx.lanl.gov/abs/1403.4545}{{\tt
  1403.4545}}.

\bibitem{Nakayama:2014lva}
Y.~Nakayama and T.~Ohtsuki, ``{Approaching conformal window of $O(n)\times
  O(m)$ symmetric Landau-Ginzburg models from conformal bootstrap},'' {\em
  Phys.Rev.} {\bf D89} (2014) 126009,
  \href{http://xxx.lanl.gov/abs/1404.0489}{{\tt 1404.0489}}.

\bibitem{Nakayama:2014yia}
Y.~Nakayama and T.~Ohtsuki, ``{Five dimensional $O(N)$-symmetric CFTs from
  conformal bootstrap},'' {\em Phys. Lett.} {\bf B734} (2014) 193--197,
  \href{http://xxx.lanl.gov/abs/1404.5201}{{\tt 1404.5201}}.

\bibitem{Alday:2014qfa}
L.~F. Alday and A.~Bissi, ``{Generalized bootstrap equations for $
  \mathcal{N}=4 $ SCFT},'' {\em JHEP} {\bf 02} (2015) 101,
  \href{http://xxx.lanl.gov/abs/1404.5864}{{\tt 1404.5864}}.

\bibitem{Chester:2014fya}
S.~M. Chester, J.~Lee, S.~S. Pufu, and R.~Yacoby, ``{The $ \mathcal{N}=8 $
  superconformal bootstrap in three dimensions},'' {\em JHEP} {\bf 1409} (2014)
  143, \href{http://xxx.lanl.gov/abs/1406.4814}{{\tt 1406.4814}}.

\bibitem{Kos:2014bka}
F.~Kos, D.~Poland, and D.~Simmons-Duffin, ``{Bootstrapping Mixed Correlators in
  the 3D Ising Model},'' {\em JHEP} {\bf 1411} (2014) 109,
  \href{http://xxx.lanl.gov/abs/1406.4858}{{\tt 1406.4858}}.

\bibitem{Caracciolo:2014cxa}
F.~Caracciolo, A.~C. Echeverri, B.~von Harling, and M.~Serone, ``{Bounds on OPE
  Coefficients in 4D Conformal Field Theories},'' {\em JHEP} {\bf 10} (2014)
  20, \href{http://xxx.lanl.gov/abs/1406.7845}{{\tt 1406.7845}}.

\bibitem{Nakayama:2014sba}
Y.~Nakayama and T.~Ohtsuki, ``{Bootstrapping phase transitions in QCD and
  frustrated spin systems},'' {\em Phys. Rev.} {\bf D91} (2015), no.~2 021901,
  \href{http://xxx.lanl.gov/abs/1407.6195}{{\tt 1407.6195}}.

\bibitem{Golden:2014oqa}
J.~Golden and M.~F. Paulos, ``{No unitary bootstrap for the fractal Ising
  model},'' {\em JHEP} {\bf 03} (2015) 167,
  \href{http://xxx.lanl.gov/abs/1411.7932}{{\tt 1411.7932}}.

\bibitem{Chester:2014mea}
S.~M. Chester, J.~Lee, S.~S. Pufu, and R.~Yacoby, ``{Exact Correlators of BPS
  Operators from the 3d Superconformal Bootstrap},'' {\em JHEP} {\bf 1503}
  (2015) 130, \href{http://xxx.lanl.gov/abs/1412.0334}{{\tt 1412.0334}}.

\bibitem{Paulos:2014vya}
M.~F. Paulos, ``{JuliBootS: a hands-on guide to the conformal bootstrap},''
  \href{http://xxx.lanl.gov/abs/1412.4127}{{\tt 1412.4127}}.

\bibitem{Beem:2014zpa}
C.~Beem, M.~Lemos, P.~Liendo, L.~Rastelli, and B.~C. van Rees, ``{The
  ${\mathcal N}=2$ superconformal bootstrap},''
  \href{http://xxx.lanl.gov/abs/1412.7541}{{\tt 1412.7541}}.

\bibitem{Simmons-Duffin:2015qma}
D.~Simmons-Duffin, ``{A Semidefinite Program Solver for the Conformal
  Bootstrap},'' {\em JHEP} {\bf 06} (2015) 174,
  \href{http://xxx.lanl.gov/abs/1502.02033}{{\tt 1502.02033}}.

\bibitem{bobev2015bootstrapping}
N.~Bobev, S.~El-Showk, D.~Mazac, and M.~F. Paulos, ``{Bootstrapping the
  Three-Dimensional Supersymmetric Ising Model},'' {\em Phys. Rev. Lett.} {\bf
  115} (2015), no.~5 051601, \href{http://xxx.lanl.gov/abs/1502.04124}{{\tt
  1502.04124}}.

\bibitem{bobev2015bootstrapping2}
N.~Bobev, S.~El-Showk, D.~Mazac, and M.~F. Paulos, ``{Bootstrapping SCFTs with
  Four Supercharges},'' {\em JHEP} {\bf 08} (2015) 142,
  \href{http://xxx.lanl.gov/abs/1503.02081}{{\tt 1503.02081}}.

\bibitem{Kos:2015mba}
F.~Kos, D.~Poland, D.~Simmons-Duffin, and A.~Vichi, ``{Bootstrapping the $O(N)$
  Archipelago},'' \href{http://xxx.lanl.gov/abs/1504.07997}{{\tt 1504.07997}}.

\bibitem{chester2015accidental}
S.~M. Chester, S.~Giombi, L.~V. Iliesiu, I.~R. Klebanov, S.~S. Pufu, and
  R.~Yacoby, ``{Accidental Symmetries and the Conformal Bootstrap},''
  \href{http://xxx.lanl.gov/abs/1507.04424}{{\tt 1507.04424}}.

\bibitem{Beem:2015aoa}
C.~Beem, M.~Lemos, L.~Rastelli, and B.~C. van Rees, ``{The $(2,0)$
  superconformal bootstrap},'' \href{http://xxx.lanl.gov/abs/1507.05637}{{\tt
  1507.05637}}.

\bibitem{iliesiu2015bootstrapping}
L.~Iliesiu, F.~Kos, D.~Poland, S.~S. Pufu, D.~Simmons-Duffin, and R.~Yacoby,
  ``{Bootstrapping 3D Fermions},''
  \href{http://xxx.lanl.gov/abs/1508.00012}{{\tt 1508.00012}}.

\bibitem{poland2015exploring}
D.~Poland and A.~Stergiou, ``{Exploring the Minimal 4D $\mathcal{N}=1$ SCFT},''
  \href{http://xxx.lanl.gov/abs/1509.06368}{{\tt 1509.06368}}.

\bibitem{Lemos:2015awa}
M.~Lemos and P.~Liendo, ``{Bootstrapping ${\mathcal N}=2$ chiral
  correlators},'' \href{http://xxx.lanl.gov/abs/1510.03866}{{\tt 1510.03866}}.

\bibitem{DO1}
F.~Dolan and H.~Osborn, ``{Conformal four point functions and the operator
  product expansion},'' {\em Nucl.Phys.} {\bf B599} (2001) 459--496,
  \href{http://xxx.lanl.gov/abs/hep-th/0011040}{{\tt hep-th/0011040}}.

\bibitem{DO2}
F.~Dolan and H.~Osborn, ``{Conformal partial waves and the operator product
  expansion},'' {\em Nucl.Phys.} {\bf B678} (2004) 491--507,
  \href{http://xxx.lanl.gov/abs/hep-th/0309180}{{\tt hep-th/0309180}}.

\bibitem{DO3}
F.~Dolan and H.~Osborn, ``{Conformal Partial Waves: Further Mathematical
  Results},'' \href{http://xxx.lanl.gov/abs/1108.6194v2}{{\tt 1108.6194v2}}.

\bibitem{Costa:2011dw}
M.~S. Costa, J.~Penedones, D.~Poland, and S.~Rychkov, ``{Spinning Conformal
  Blocks},'' {\em JHEP} {\bf 1111} (2011) 154,
  \href{http://xxx.lanl.gov/abs/1109.6321}{{\tt 1109.6321}}.

\bibitem{echeverri2015deconstructing}
A.~C. Echeverri, E.~Elkhidir, D.~Karateev, and M.~Serone, ``{Deconstructing
  Conformal Blocks in 4D CFT},'' {\em JHEP} {\bf 08} (2015) 101,
  \href{http://xxx.lanl.gov/abs/1505.03750}{{\tt 1505.03750}}.

\bibitem{SimmonsDuffin:2012uy}
D.~Simmons-Duffin, ``{Projectors, Shadows, and Conformal Blocks},'' {\em JHEP}
  {\bf 1404} (2014) 146, \href{http://xxx.lanl.gov/abs/1204.3894}{{\tt
  1204.3894}}.

\bibitem{Costa:2014rya}
M.~S. Costa and T.~Hansen, ``{Conformal correlators of mixed-symmetry
  tensors},'' {\em JHEP} {\bf 1502} (2015) 151,
  \href{http://xxx.lanl.gov/abs/1411.7351}{{\tt 1411.7351}}.

\bibitem{rejon2015scalar}
F.~Rejon-Barrera and D.~Robbins, ``{Scalar-Vector Bootstrap},''
  \href{http://xxx.lanl.gov/abs/1508.02676}{{\tt 1508.02676}}.

\bibitem{Hogervorst:2013sma}
M.~Hogervorst and S.~Rychkov, ``{Radial Coordinates for Conformal Blocks},''
  {\em Phys.Rev.} {\bf D87} (2013), no.~10 106004,
  \href{http://xxx.lanl.gov/abs/1303.1111}{{\tt 1303.1111}}.

\bibitem{penedones2015recursion}
J.~Penedones, E.~Trevisani, and M.~Yamazaki, ``{Recursion Relations for
  Conformal Blocks},'' \href{http://xxx.lanl.gov/abs/1509.00428}{{\tt
  1509.00428}}.

\end{thebibliography}\endgroup
\bibliographystyle{ssg}

\end{document}